\documentclass[preprint,amsmath,amssymb,superscriptaddress,aip,jap,floatfix,showpacs]{revtex4-1}

\usepackage{graphicx}
\usepackage{bm}
\usepackage{amsmath}

\begin{document}

\begin{flushleft} The following article has been accepted by Journal of Applied Physics. After it is published, it will be found at https://publishing.aip.org/resources/librarians/products/journals/ \end{flushleft}
\title{Enhancement of the electronic thermoelectric properties of bulk strained silicon-germanium alloys using the scattering relaxation times from first principles}

\author{F.~Murphy-Armando}
\affiliation{Tyndall National Institute, University College Cork, Lee Maltings, Dyke Parade, Cork, T12 R5CP, Ireland}
\email{philip.murphy@tyndall.ie}

\date{\today{}}

\begin{abstract}
We use first-principles electronic structure methods to calculate the electronic thermoelectric properties (i.e. due to electronic transport only) of single-crystalline bulk $n$-type silicon-germanium alloys vs Ge composition, temperature, doping concentration and strain. We find excellent agreement to available experiments for the resistivity, mobility and Seebeck coefficient. These results are combined with the experimental lattice thermal conductivity to calculate the thermoelectric figure of merit $ZT$, finding very good agreement with experiment. We predict that  3\% tensile hydrostatic strain enhances the $n$-type $ZT$ by 50\% at carrier concentrations of $n=10^{20}$ cm$^{-3}$ and temperature of $T=1200K$. These enhancements occur at different alloy compositions due to different effects: at 50\% Ge composition the enhancements are achieved by a strain induced decrease in the Lorenz number, while the power factor remains unchanged. These characteristics are important for highly doped and high temperature materials, in which up to 50\% of the heat is carried by electrons. At 70\% Ge the increase in $ZT$ is due to a large increase in electrical conductivity produced by populating the high mobility $\Gamma$ conduction band valley, lowered in energy by strain. 


\end{abstract}


\maketitle
\section{Introduction}

Improving the performance of thermoelectric materials is hard. 
The main reason for this is that the efficiency depends on several inter-related properties of the material which are not possible to optimise individually. The efficiency is determined by the dimensionless product $ZT$, referred to as the figure of merit of the material. The figure of merit is a function of the electronic and lattice properties of the material\cite{ioffe}:
\begin{equation}\label{eqzt}
ZT=\frac{T\sigma S^2}{\kappa_e+\kappa_{ph}},
\end{equation}
where $S$ is the Seebeck coefficient, $\sigma$ the carrier conductivity, and $\kappa_e$ and $\kappa_{ph}$ are the carrier and lattice thermal conductivities, respectively.

The quantities $S$, $\sigma$ and $\kappa_e$ are determined by the electronic properties of the material, viz., the electronic band structure and carrier scattering mechanisms, while $\kappa_{ph}$ is a function of the phonon dispersion and lifetimes. All these quantities are interrelated (the latter, $\kappa_{ph}$, is related directly by electron-phonon scattering of the phonons, and indirectly by any measures taken to decrease the lattice thermal conductivity that has an effect on the electronic properties and vice versa). It is therefore very challenging to increase the figure of merit, as steps taken to enhance one of these quantities may adversely affect the other variables. The game is then one of fine tuning the material properties to find a simultaneous optimum in all these quantities.

Several strategies have been found that yield improvements in $ZT$.\cite{advth} Historically, most have relied on reducing $k_{ph}$, as it is phonon dependent and mostly decoupled from the other variables. Regarding the electronic properties, promising strategies include band convergence\cite{bconv1,bconv2} and mobility enhancements, among others\cite{advth}.

In this work we use {\it ab initio} electronic structure methods to explore two unusual ways of improving the thermoelectric efficiency of a highly doped $n$-type single crystalline SiGe semiconductor alloy by applying strain: 
by reducing the Lorenz number while avoiding a decrease in power factor, and by increasing the mobility while minimising a decrease in Seebeck coefficient. Both cases illustrate different aspects of band convergence and mobility enhancement. We also compare our results for the unstrained case to available experiments.

Finding a maximum in $ZT$ is akin to charting the highest peaks in a misty, multidimensional landscape of temperature, material compositions, strain, doping concentration and nano-structural features. The large number of variables makes this a very costly problem to solve solely experimentally. It is still a daunting task computationally, but high-throughput techniques are being developed to speed up the process\cite{ht1,ht2}. These techniques rely on relatively fast calculations of the electronic and phonon band structure of combinations of materials, usually treating the electronic scattering in a simplified way. These methods provide good intuitions and point towards good candidate materials, narrowing the number of materials on which full calculations and experiments need to be made. However, electronic scattering is a very strong determinant of the overall $ZT$, that varies substantially with material characteristics. Not having reasonably accurate scattering rates may lead to both false positives and overlooking very strong candidates. Scattering rates, however, are hard to measure and very time consuming to calculate, to be introduced directly into an automated search for high $ZT$.

Many works have addressed the different aspects of this problem using first principles calculations. Most have calculated the lattice thermal conductivity from first principles, a few have used first principles band structures to calculate the electronic properties, and fewer still have calculated the scattering rates\cite{fpcal1,fpcal2,fpcal3,fpcal4,fpcal5,fpcal6,fpcal7,mingosi,jiangmethod,prl1,prb}. 
We use \textit{ab initio} derived band structures and our previously calculated scattering parameters\cite{prl1,prb,prb2,jap2,jap} for all SiGe compositions and hydrostatic strain configurations, also from first principles. We include the effects of temperature on the band structure, including the non-parabolicity of the bands, and find it has a noticeable effect on the thermoelectric properties.

The thermoelectric properties are calculated using the Boltzmann transport equation in the (momentum dependent) relaxation time approximation. The electrical conductivity in terms of the scattering processes is given by\cite{ashcroft}
\begin{equation}
{\bf \sigma}=\sum_i{\bf L}_i^{11},
\end{equation}
the Seebeck coefficient by
\begin{equation}
\label{seebeck}
{\bf S}=\frac{\sum_i{\bf L}_i^{12}}{{\bf \sigma}},
\end{equation}
the electronic thermal conductivity by
\begin{equation}
{\bf \kappa_e}=\sum_i{\bf L}_i^{22}-\sum_i{\bf L}_i^{21}{\bf S},
\end{equation}
and the Lorenz number $L$ by
\begin{equation}\label{kappa}
L=\frac{\bf \kappa_e}{{\bf \sigma}T}
\end{equation}
where the sum is over all occupied bands (conduction and valence) and
\begin{eqnarray}
{\bf L^{11}}&=&{\bf \mathcal{L}}^{(0)}, \nonumber\\
{\bf L^{21}}&=&T{\bf L^{12}}=-\frac{1}{e}{\bf \mathcal{L}}^{(1)}, \nonumber\\
{\bf L^{22}}&=&\frac{1}{e^2T}{\bf\mathcal{L}}^{(2)},
\end{eqnarray}
and the scattering kernel ${\bf\mathcal{L}}^{(\alpha)}$ is given by
\begin{equation}\label{kernel}
{\bf\mathcal{L}}^{(\alpha)}=
 e^2\int \frac{d{\bf k}}{4 \pi^3}\left(-\frac{\partial f}{\partial \varepsilon}\right)\tau_{\bf k} \bf{v}({\bf k})\bf{v}({\bf k}) \left(\varepsilon({\bf k})-\mu\right)^\alpha.
\end{equation}
Here, $\tau_{\bf k}$ is the $k$-dependent relaxation time, $f$ the electronic distribution, $\bf v$ the group velocity, $\varepsilon$ the electronic energy and $\mu$ the Fermi Level. 
\section{Method}\label{method}
In this work we use electron-phonon and alloy scattering parameters calculated previously from first principles. We integrate Eq. \ref{kernel} with the band structure of the strained SiGe alloy calculated analytically using the {\bf{k.p}} approach of Rideau {\it{et al}}.\cite{kdp} This approach is based on empirical parameters and first principles GW calculations, and validated against GW calculations. We corrected the resulting band edges to include the alloy-induced shifts using the Coherent Potential Approximation (CPA) as in Ref. \onlinecite{prl1}. While the {\bf{k.p}} approach of Ref. \onlinecite{kdp} is not fully first-principles, its analytic form and dependence in strain gives a very versatile tool to search the wide parameter space needed in this work. The results given by this {\bf{k.p}} approach vs using the band structure from DFT and GW calculations are almost identical to our previous results for alloy and phonon-limited transport.\cite{prb2,prberratum,jap2,jap} Also, other groups are following similar {\bf{k.p}} approaches to interpolate the DFT generated band structure\cite{berland1} to ease the integration of thermoelectric properties.\cite{berland2}

The integration of Eq. \ref{kernel} is made much easier if carried out in energy, rather than momentum. The $\Delta$ and $L$ valleys have a near parabolic dispersion that allows parametrization in energy. For these two valleys, assuming a first-order non-parabolic dispersion we get the relation 
\begin{equation}
\varepsilon(1+\alpha \varepsilon)=\frac{\hbar^2 k^2}{2m^*},
\end{equation}
where $k$ the electron crystal momentum and $\alpha$ the non-parabolicity factor. On the other hand, the $\Gamma$ valley is highly non-parabolic, and is integrated in $k$-space.

The expressions of the group velocity and the density of states mass in eq. \ref{kernel} become\cite{jacob}
\begin{equation}
v({\bf k})=\frac{1}{\hbar}\frac{d\varepsilon}{d{\bf k}}=\frac{\hbar {\bf k}}{m (1+2\alpha)\varepsilon},
\end{equation}
and
\begin{equation}
m_d^{\frac{3}{2}}=\sqrt{\frac{m_t^2 m_l}{\pi \alpha k_B T}}e^{\frac{1}{2 \alpha k_B T}} K_2 (2 \alpha k_B T),
\end{equation}
where $k_B$ is the Boltzmann constant, $K_2$ is the modified Bessel function of order 2, $m_t$ and $m_l$ the transverse and longitudinal effective masses of the valley in question, respectively. The modification by non-parabolicity of the usual parabolic expressions of the alloy, electron-phonon and impurity scattering are treated in Ref. \onlinecite{jacob}.

The temperature dependence of the electronic band structure is included via the dependence of the valley band edges and non-parabolicity factor. The $\Delta$, $L$ and $\Gamma$ energies as a function of temperature are taken from Ref. \onlinecite{varshni}, while the non-parabolicity factor was calculated as explained below.

\begin{figure}
\includegraphics[width=3.4in]{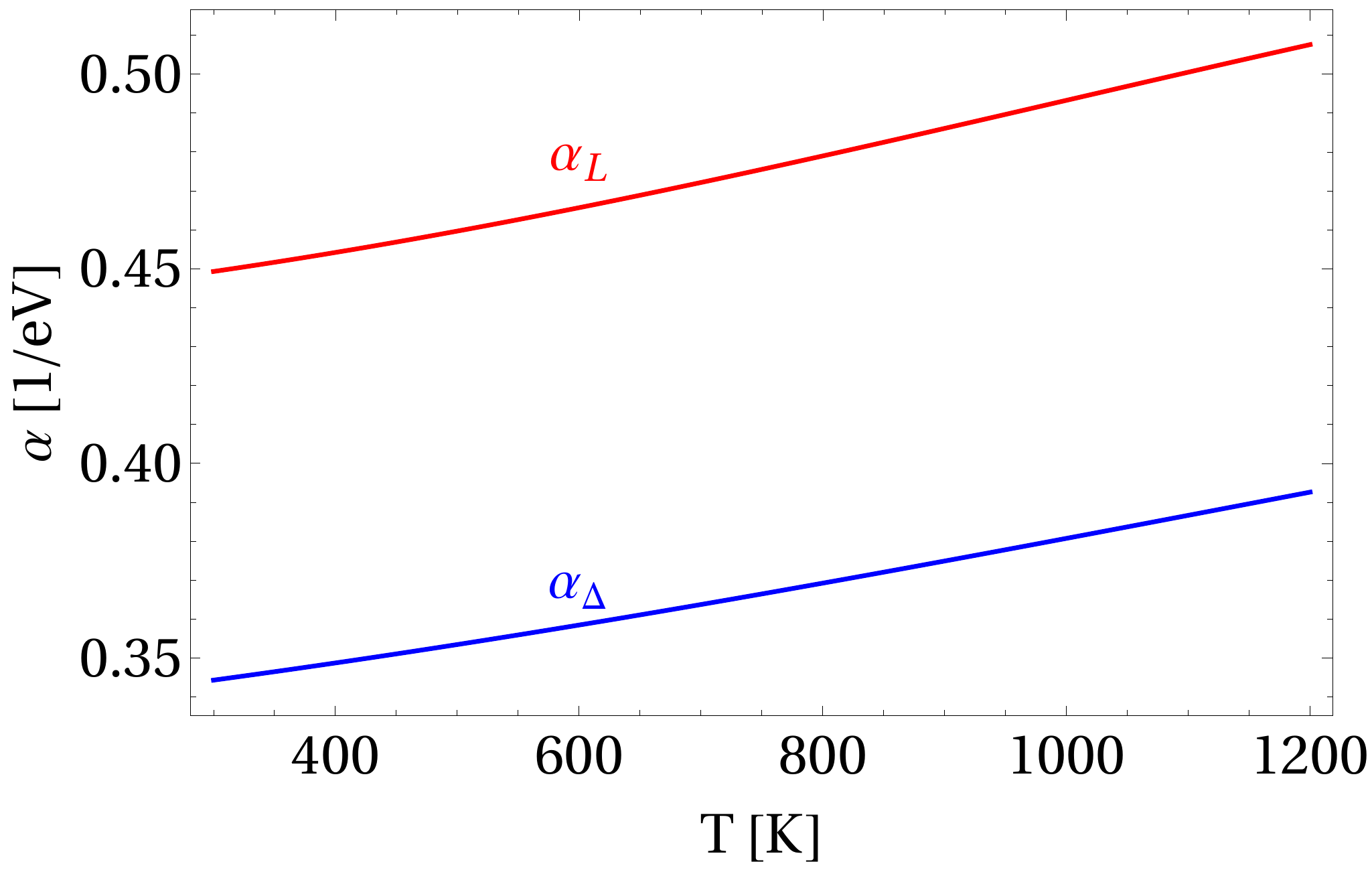}
\caption{\label{alphap} Non-parabolicity factor $\alpha$ vs temperature for the $\Delta$ and $L$ conduction band valleys.}
\end{figure}

\begin{figure}
\includegraphics[width=3.4in]{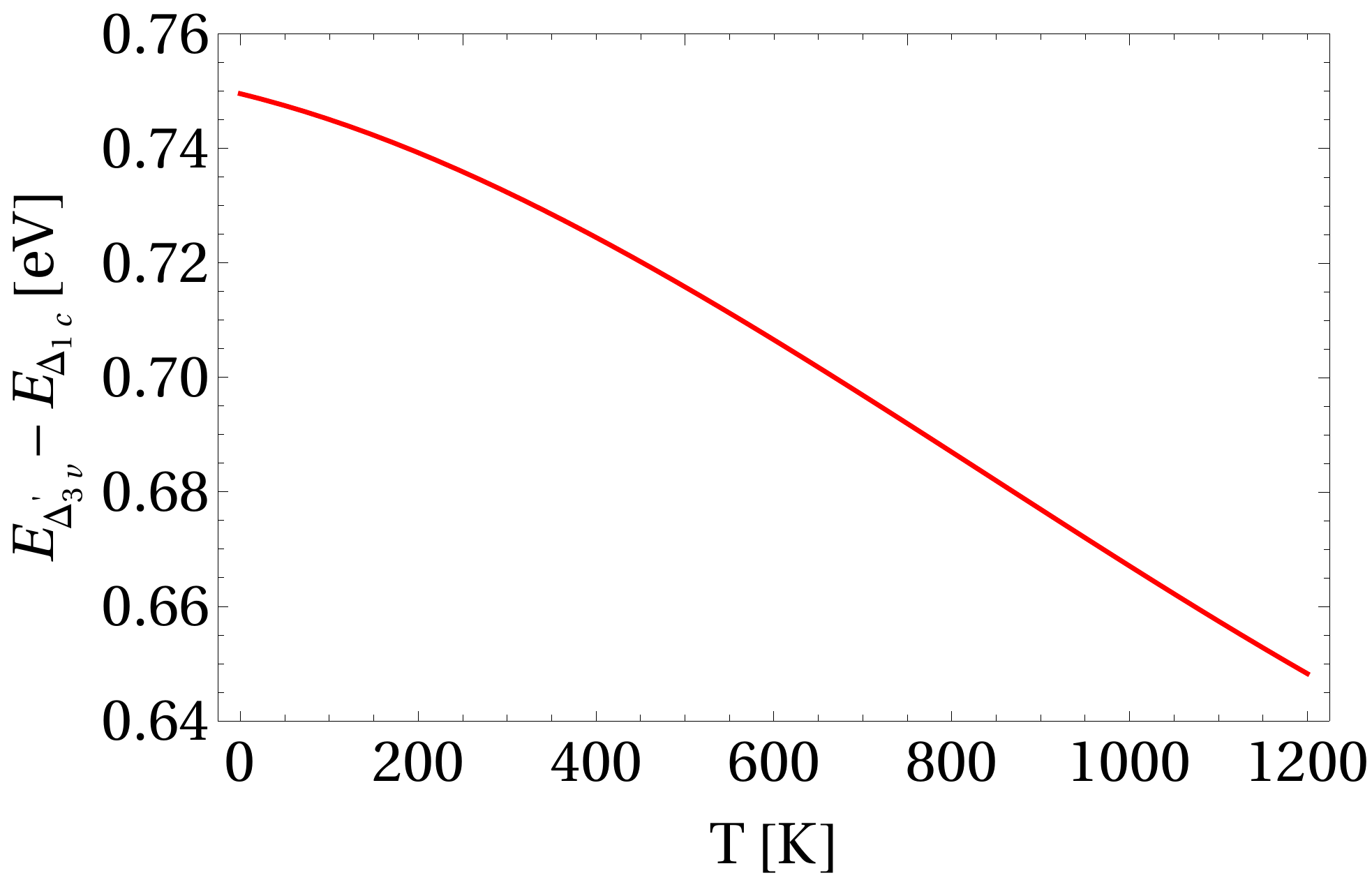}
\caption{\label{dedelta} Energy gap between the $E_{\Delta^{'}_{3v}}$ and $E_{\Delta_{1c}}$ energies at the $\Delta$ point vs temperature.}
\end{figure}

As we will show later, we found that the temperature dependence of the non-parabolic effects in the carrier energies are important. Non-parabolicity is determined by the proximity of other bands to the energy band populated by the carriers. 
Just like the band gap of the material, the band energy differences that give rise to non-parabolicity are temperature dependent, resulting in a temperature dependent non-parabolicity factor.\cite{sigemodelling}
Here, we show for the first time the energy dependence of bands that affect non-parabolicity in SiGe. They are calculated from first-principles using the recent methods based on density functional perturbation theory based on the Allen-Heine-Cardona approach,\cite{Allen1976,Allen1981,Allen1983} available in Abinit implemented by Samuel {\it et al}.\cite{ponce1,ponce2} The expressions for the non-parabolicity factor $\alpha$ \cite{jacob} for the $L$\cite{paige} and $\Delta$\cite{jacob} bands in SiGe are given by
\begin{eqnarray}
\alpha_L&=&\frac{1}{E_{L'_{3v}}-E_{L_{1c}}},\\
\alpha_{\Delta}&=&\frac{1}{2\left(E_{\Delta^{'}_{3v}}-E_{\Delta_{1c}}\right)}\left(1-\frac{m_t}{m_e}\right)^2,
\end{eqnarray}
where $E_{i_j}$ are the band energies with $j$ symmetry at the corresponding band minimum $i$, $m_t$ the effective mass of the $\Delta$  conduction band valley in the transverse direction and $m_e$ the electron mass. The resulting factors as a function of temperature are shown in Fig. \ref{alphap}. 
The temperature dependence of the band gap $E_{L'_{3v}}-E_{L_{1c}}$  is available in the literature from experiment\cite{varshni}. 
The band gap $E_{\Delta^{'}_{3v}}-E_{\Delta_{1c}}$ is not available in the literature, and we have calculated it using the Samuel's approach, shown in Fig. \ref{dedelta}.
The non-parabolicity of the $\Gamma$ band cannot be described by a single factor, and is considered directly through integration in $k$-space, rather than energy, for all quantities. While we do consider the effects of temperature of $\Gamma$ band edge energy, we ignore these effects on the band dispersion. This results in a more conservative estimate of the increase in power factor, as increasing temperature decreases the $\Gamma$ effective mass. In the cases considered in this work, the $\Gamma$ band dispersion is affected more by strain than temperature, and are included in the calculations.

The changes in electronic energy with temperature in this work are taken for the unstrained case. Strain may affect the energy response to temperature, but we expect it to be small relative to the strain response.

Most of the scattering parameters used in this work can be found in the literature. In our previous work, we calculated from first principles the scattering by alloy disorder\cite{prl1,prb} and electron-phonon scattering\cite{prb2,prberratum}. We also calculated the effect of strain in these parameters. Strain affects mostly the scattering by acoustic phonons,\cite{jap2,jap} while it has little effect on inter-valley scattering and scattering by optical phonons.\cite{jap2,jelena} Here we are chiefly concerned with $n$-type carrier thermoelectric properties. However, at high temperatures some $p$-type transport needs to be considered to fully account for the bi-polar behaviour. We take the parameters for $p$-type transport from the literature: the hole-acoustic deformation potentials are from Ref. \onlinecite{kdp}, while the formulae for the hole mobility and and optical phonon scattering are from the approach of Fischetti and Laux\citep{fisch}.
The ionized impurity scattering for $n$-type SiGe is included using the Brooks-Herring approach.\cite{jacob,chatt,barrie}

\begin{figure}
\includegraphics[width=3.4in]{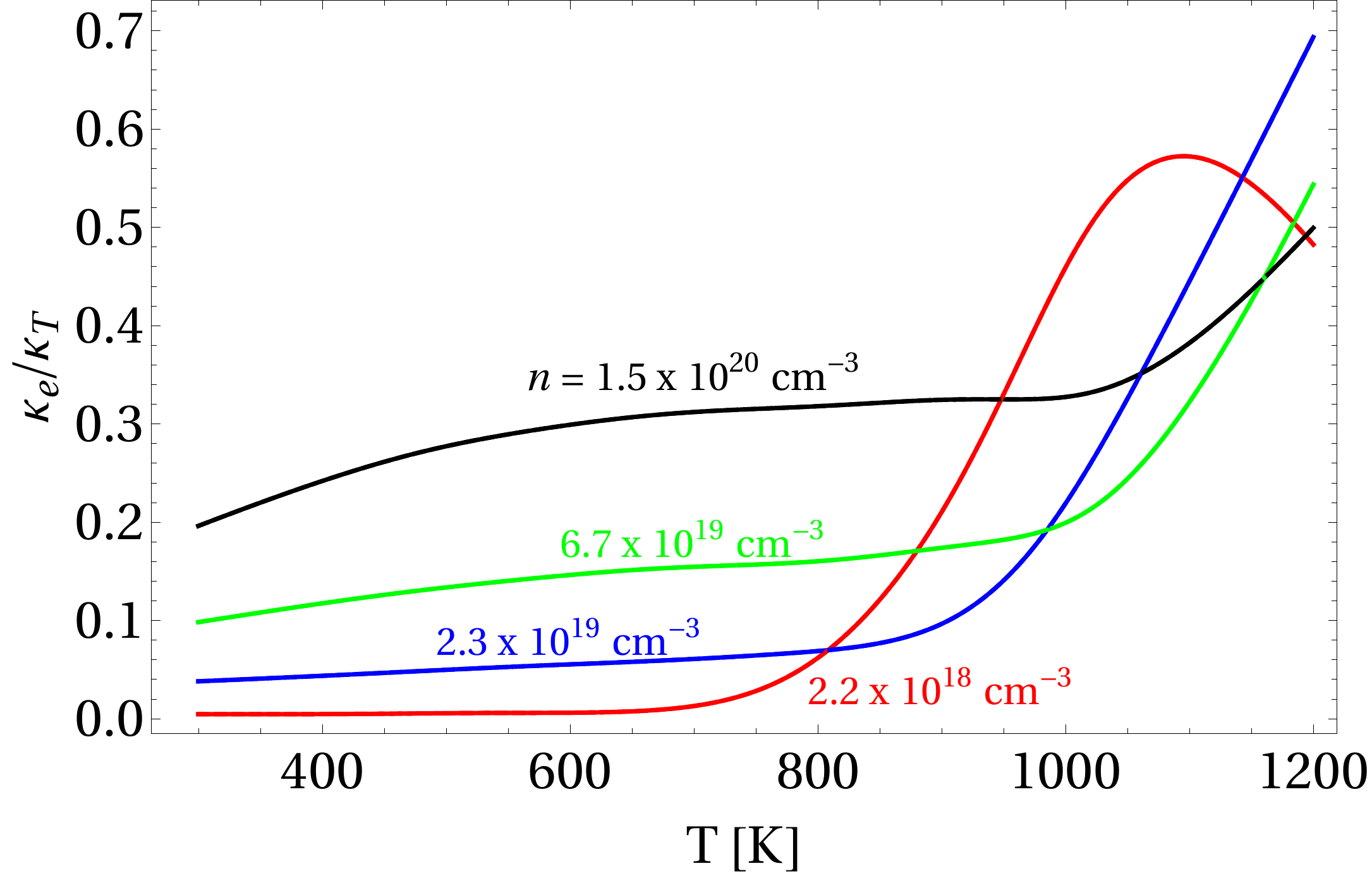}
\caption{\label{kratio} Ratio $\kappa_e/\left(\kappa_e+\kappa_{ph}\right)$ of the electronic to total thermal conductivity vs temperature for four doping concentrations. $\kappa_e$ is calculated using Eq. \ref{kappa}.}
\end{figure}

We treat the effect of strain on the thermal conductivity only via the electronic contribution.  We expect the changes in lattice thermal conductivity to be of the order of 1\% per percent strain. In this work we do not calculate the lattice thermal conductivity $\kappa_{ph}=\kappa-\kappa_e$, and we obtain it from a fit to measured values of $\kappa$ of Ref. \onlinecite{dismukes}, subtracting our calculated $\kappa_e$. While calculations of this quantity are possible using first principles calculations, they are not available in SiGe for the whole range of doping concentrations and temperatures, and they generally do not include electron-phonon scattering. To include the effects of strain on the total $\kappa$, we have assumed that the lattice thermal conductivity is not affected by strain, and added the calculated change in electronic thermal conductivity. The proportion of the electronic to total thermal conductivity as a function of temperature for several doping concentrations is shown in Fig. \ref{kratio}, as calculated using this approach and Eq. \ref{kappa}. We see that at high temperatures and doping concentrations the electrons can carry as much heat at the lattice.


\section{Results}

\subsection{Comparison to Experiment}

\begin{figure}
\includegraphics[width=3.4in]{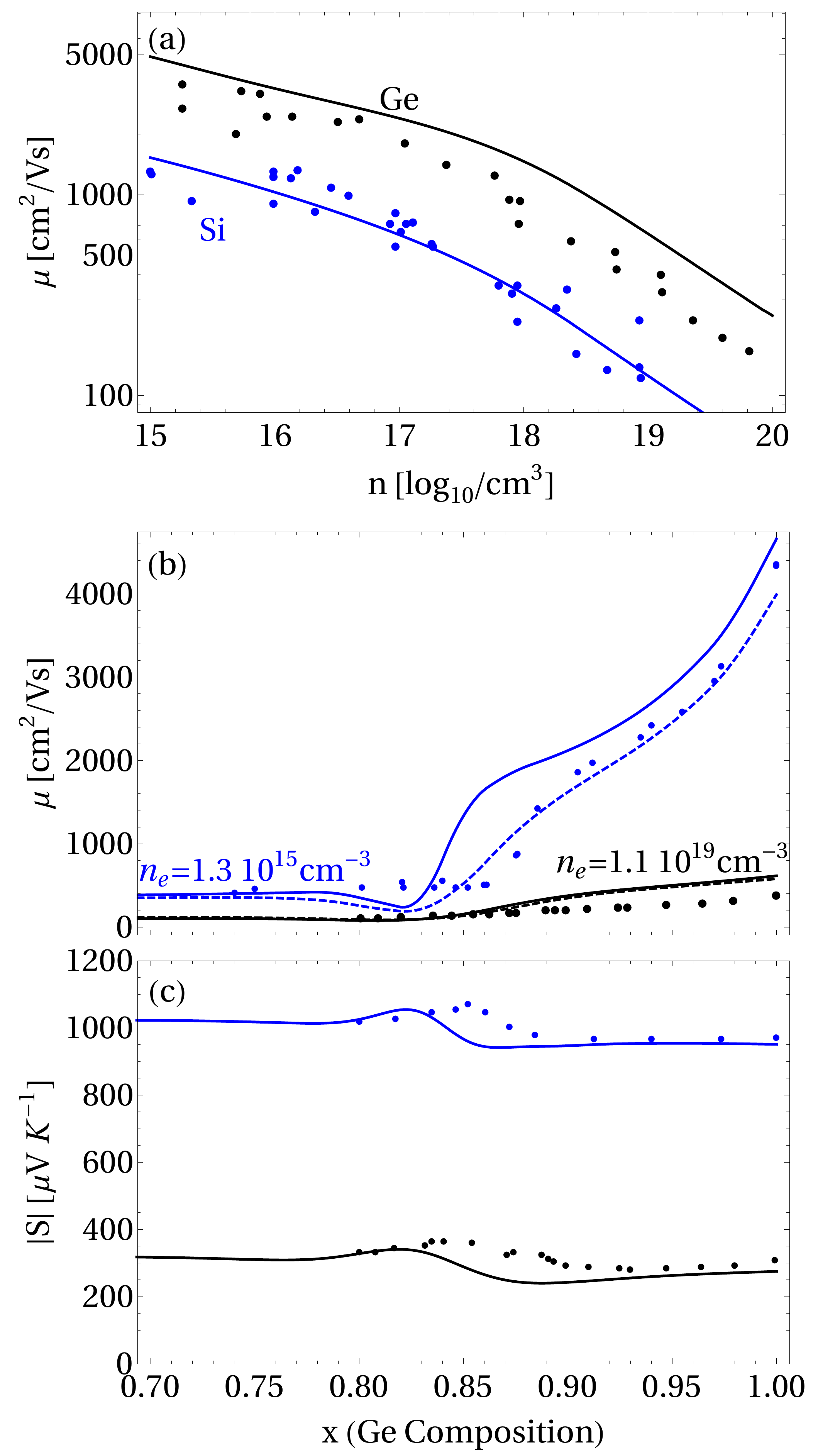}
\caption{\label{muandseevnvx} Theoretical and experimental comparison of the room temperature (a) mobility $\mu$ of Si and Ge vs doping concentration $n$. The dots are the experimental values of Ref. \onlinecite{expge} for Ge and Refs. \onlinecite{expsi2}, \onlinecite{expsi3} and \onlinecite{expsi4}. (b) The Hall (dashed line) and drift (solid line) mobilities $\mu$ and (c) Seebeck coefficient $S$ vs Ge concentration $x$ for high (black) and low (blue) doping concentration. Dots show the experimental values of the Hall mobility and Seebeck coefficient of Ref. \onlinecite{amith}}
\end{figure}

\begin{figure*}
\includegraphics[width=6.8in]{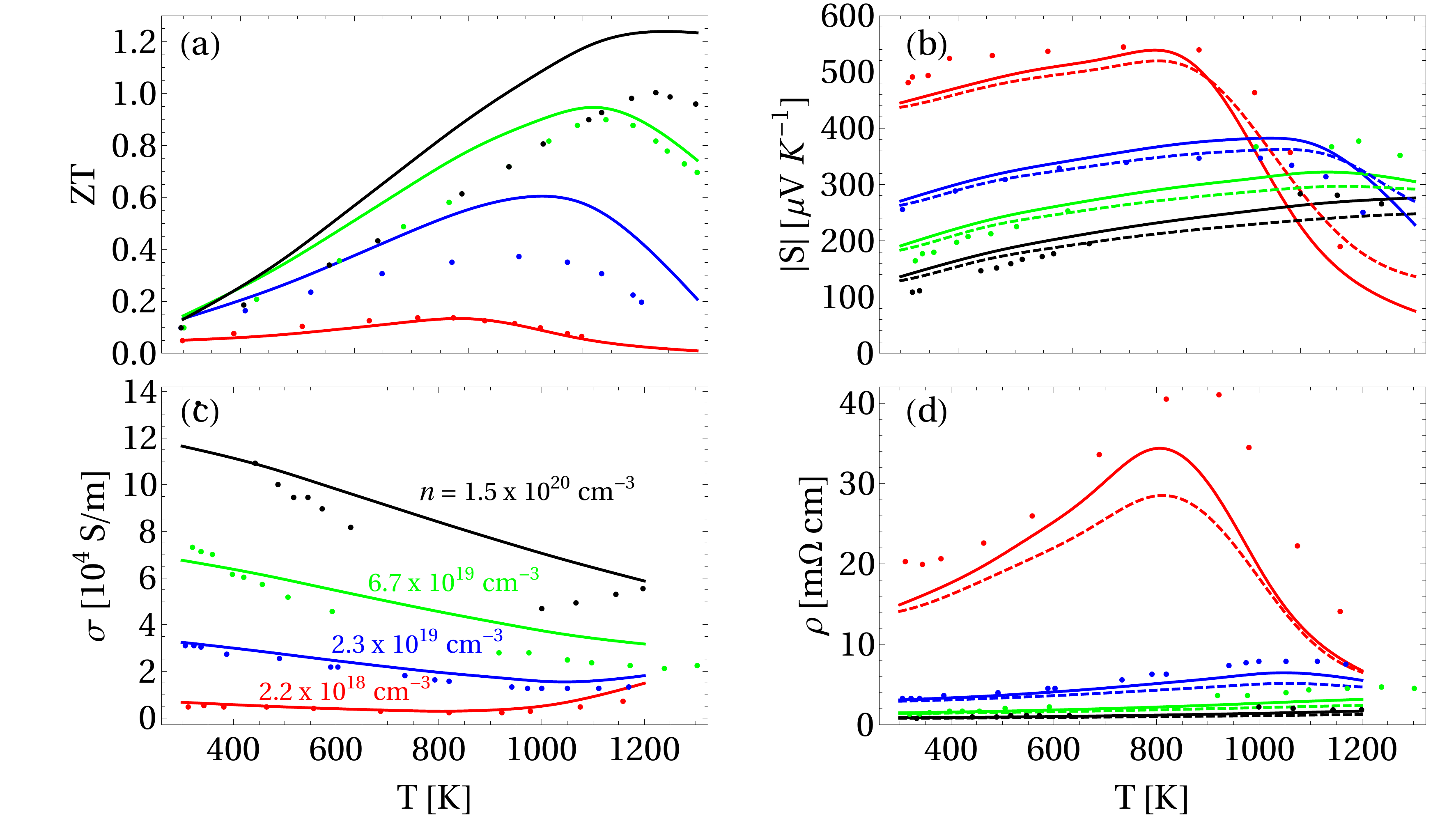}
\caption{\label{theoexpx3} Theoretical and experimental thermolectric properties of a 30\% Ge alloy vs. temperature $T$ for various doping concentrations. 
The panels show (a) thermoelectric $ZT$, (b) Seebeck coefficient $S$, (c) electrical conductivity $\sigma$ and (d) electrical resistivity $\rho$. Dashed lines show the effects of not including the temperature dependence in the non-parabolicity of the bands.}
\end{figure*}

\begin{figure}
\includegraphics[width=3.4in]{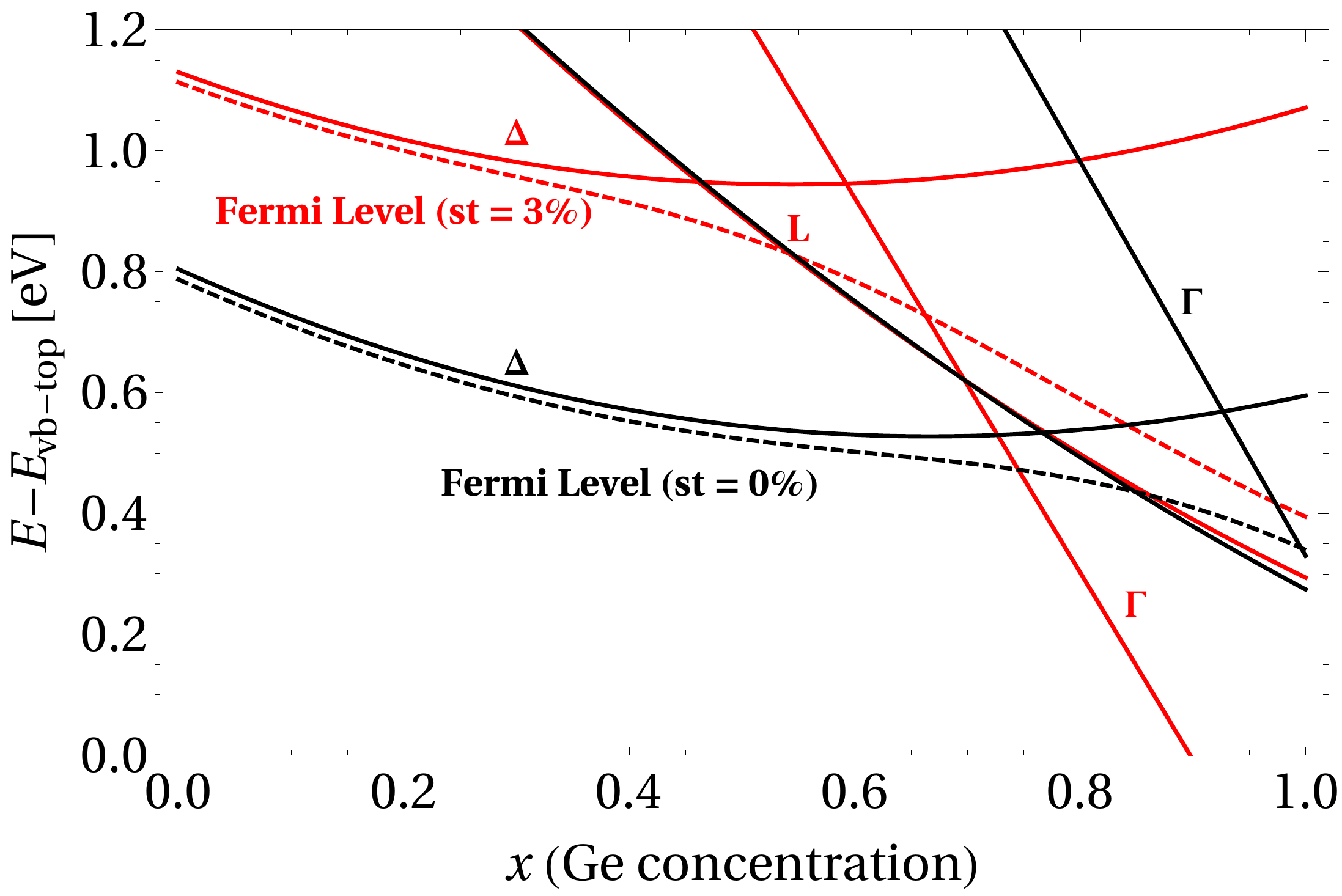}
\caption{\label{bandsn1p5d20} Energy of the bottom of the $L$, $\Gamma$ and $\Delta$ conduction band valleys vs Ge concentration $x$ at $T$=1200K and $n$=$1.5\times 10^{19}$cm$^{-3}$, for unstrained (st = 0\%, black) and st = 3\% tensile hydrostatic strained (red) Si$_{1-x}$Ge$_x$. The line corresponding to each valley is labelled on each line. The energy of the $L$ valley does not change with strain, so both strain and unstrained $L$ valley are represented by one line. The Fermi level is represented by a dashed line. The Fermi level is obtained using the k.p derived bands and the resulting carrier density for each doping concentration.}
\end{figure}

In this section we compare our calculated results of the thermelectric properties of $n$-type SiGe to available experimental data. Figs. \ref{muandseevnvx} and \ref{theoexpx3} show the mobility, electrical conductivity and resitivity, Seebeck and figure of merit $ZT$ for various temperatures and doping and alloy concentrations, compared to available experiments.
Starting with the Seebeck coefficient $S$, we compare it to the experimental work of Amith\cite{amith} and Dismukes\cite{dismukes}, at various alloy compositions, doping concentrations and temperatures. Fig \ref{muandseevnvx} (c) shows the Seebeck coefficient vs Ge composition $x$ at $n$-type doping concentrations of (upper-blue) $1.3 \times 10^{15} cm^{-3}$ and (lower-black) $1.1\times 10^{19} cm^{-3}$, compared to the experimental work of Amith. The peak-like feature is due to the increase of the density of states at the $\Delta-L$ band crossover, which lowers the Fermi level relative to the conduction band and hence increases the magnitude of the Seebeck coefficient (see the st = 0\% strain case in Fig. \ref{bandsn1p5d20} for the Fermi level and relative energies of the $\Delta$ and $L$ bands vs $x$).

Fig. \ref{theoexpx3} (b) displays the calculated Seebeck coefficient vs temperature T at various doping concentrations at Ge concentration $x \sim 0.3$ (notice that $x$ is slightly different for each curve, as expressed in the source of the experimental data, Ref. \onlinecite{dismukes}. We have used the corresponding $x$ concentrations in our calculations.) The sharp drop at higher temperatures is due to the onset of intrinsic behaviour and is caused by two interrelated effects: i) as the temperature rises, the electronic Fermi distribution $f$ becomes wider in energy, resulting in the lowering of the Fermi level and increasing the amplitude of $S$. ii) As the Fermi level drops towards mid-gap and the population of holes in the valence band grows due to the broad distribution, the $p$-type contribution to the total Seebeck coefficient increases. The latter is of opposite sign to the $n$-type contribution, thus sharply decreasing the magnitude of the total $S$. At higher doping concentration, the larger number of electrons in the conduction band raises the temperature at which the intrinsic behaviour occurs. The measured higher Seebeck coefficients at higher temperatures at concentrations $n=6.7\times 10^{19} cm^{-3}$ and $1.5\times 10^{20} cm^{-3}$ are attributed to an increase in carrier concentration due to the dissolution of precipitates.\cite{dismukes} However, we would expect the behaviour of $S$ and $\rho$ (see Fig. \ref{theoexpx3} (d)) for the experiments to correspond rather to a $decrease$ in carrier concentration.

The electrical conductivity and resistivity as a function of temperature at a Ge content of $x\sim 0.3$ and various $n$-type doping concentrations are shown in Figs. \ref{theoexpx3} (c) and (d), respectively, and compared to experimental results.\cite{dismukes} We first observe that the calculated resistivities are consistently lower than the measured ones. This is because the experimental samples of Ref. \onlinecite{dismukes} are poly-crystalline, while we have assumed a perfectly crystalline solid. The presence of crystalline boundaries in the experimental samples is an additional source of scattering, which results in a higher resistivity. On the contrary, our calculations are in excellent agreement with the mobility measured by Amith\cite{amith} and Glicksmann\cite{glicks} in crystalline samples, as previously published\cite{prb2,prberratum} for low doping, and in Fig. \ref{muandseevnvx} (b) for high doping concentration. The effect of scattering on $S$ is much smaller, and hence the calculations agree better with the measured results.
At high temperatures, as in the case of the Seebeck coefficient, the resistivity drops as the number of minority carriers increases in the valence band. In Fig. \ref{muandseevnvx} (a) we show the room temperature $n$-type mobility of Si and Ge vs. doping concentration. As in Fig. \ref{muandseevnvx} (b), our results overestimate the mobilities of Ge at high doping concentration. We believe this is due to an inadequate treatment of the dielectric screening of the ionized impurities in the Brooks-Herring approach.\cite{dielectric}

Finally, Fig. \ref{theoexpx3} (a) shows the calculated and measured figure of merit $ZT$ vs. $T$ for the same samples as Figs. \ref{theoexpx3} (b), (c) and (d). The calculated figures of merit are consistently higher than those measured. This is because the calculated electronic properties consider a crystalline material, while the experiment is performed on poly-crystalline samples.\cite{dismukes}. The crystal boundaries add sources of scattering that are not considered in the model for the electronic properties. Furthermore, as explained in section \ref{method}, the thermal conductivity is fitted to the measured values of Ref. \onlinecite{dismukes}, and therefore includes the added scattering in the lattice thermal conductivity included in the model. In the experimental figure of merit, these two effects counterbalance each other (see Eq. \ref{eqzt}). However, the calculated result is a combination of the fitted lattice thermal conductivity and the calculated electrical conductivity, compounding rather than cancelling the effects of inter-domain scattering, resulting in higher $ZT$s than the measured case.

The effects of not including the temperature dependence of the non-parabolic factor $\alpha$ are shown in dashed lines in Figs. \ref{theoexpx3}(b) and (c). In general, the temperature dependence of $\alpha$ improves agreement with experiment.

\subsection{Improving $ZT$ with strain}
In this section we quantitatively explore the effects of strain on the thermoelectric properties of SiGe alloys. Interest in strained SiGe is due to its importance in enhancing the mobility in transistor channels and its use as sensitive piezo-resistive sensors.\cite{barlian,piezoprb} Regarding thermoelectric effects, strain has the potential to increase the figure of merit, via the enhancement of the electrical mobility. Strain also strongly affects the thermopower, often counterbalancing the effects on the mobility.\citep{hinsche1,hinsche2} In heterostructures, where layered materials are used to decrease the lattice thermal conductivity, the resulting inter-layer strain may have undesirable effects on the power factor.\citep{hinsche1,hinsche2}.

\begin{figure*}
\includegraphics[width=6.8in]{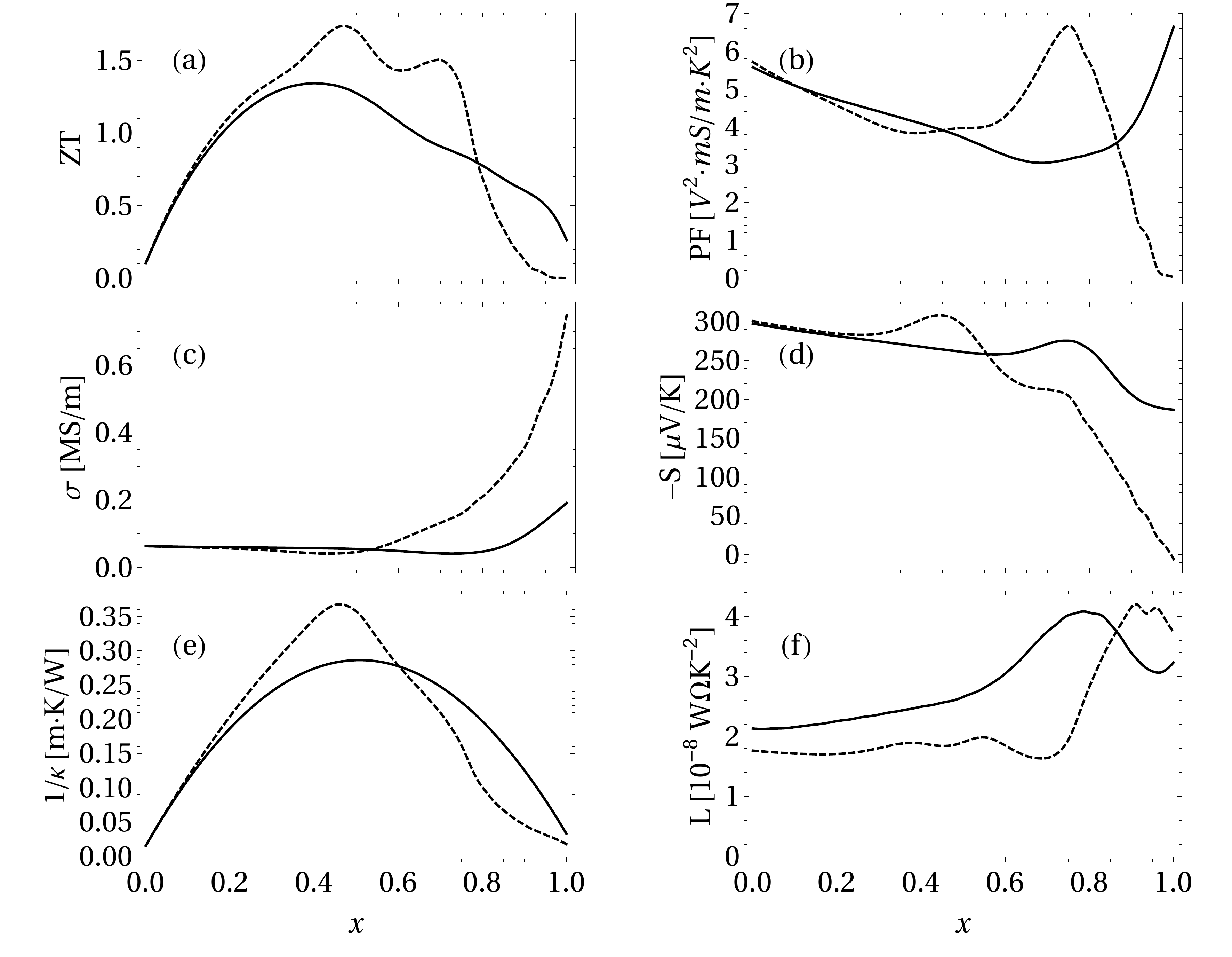}
\caption{\label{ztpfsigsiklvsxst} Si$_{1-x}$Ge$_x$ $n$-type thermoelectric properties vs Ge composition $x$ at $T=1200$K and $n=10^{20}$cm$^{-3}$. The panels show (a) thermoelectric $ZT$, (b) Power Factor PF, (c) electrical conductivity $\sigma$, (d) Seebeck coefficient $S$, (e) total inverse thermal conductivity $\kappa^{-1}=\left(\kappa_{ph}+\kappa_e\right)^{-1}$ and (f) Lorenz number $L$. Solid and dashed lines represent unstrained and 3\% hydrostatically tensile strained material.}
\end{figure*}

In many-valley semiconductors, such as SiGe, the introduction of strain typically shifts the relative energy of the valleys and redistributes the charge carrier concentrations in each valley. If such valleys are anisotropic, i.e. their effective mass is a function of crystal momentum direction, it is possible to effect a higher carrier conductivity under certain strain configurations. This requires that the direction of transport is along the lower effective mass of one or more energy valleys, and that carriers are predominantly in those valleys. As this effect depends on the proportion of carriers in the low effective mass valleys relative to others, it depends on the energy separation that can be achieved with strain, and will diminish at higher temperatures or doping concentrations. The electronic thermal conductivity $\kappa_e$, via the Wiedemann-Franz law, is affected in a similar way as the electronic conductivity $\sigma$. In materials with a high proportion of the thermal conductance being carried by electrons, such as the highly doped case, strain that increases the carrier mobility will adversely affect the figure of merit by increasing the thermal conductivity.

\begin{figure}
\includegraphics[width=3.4in]{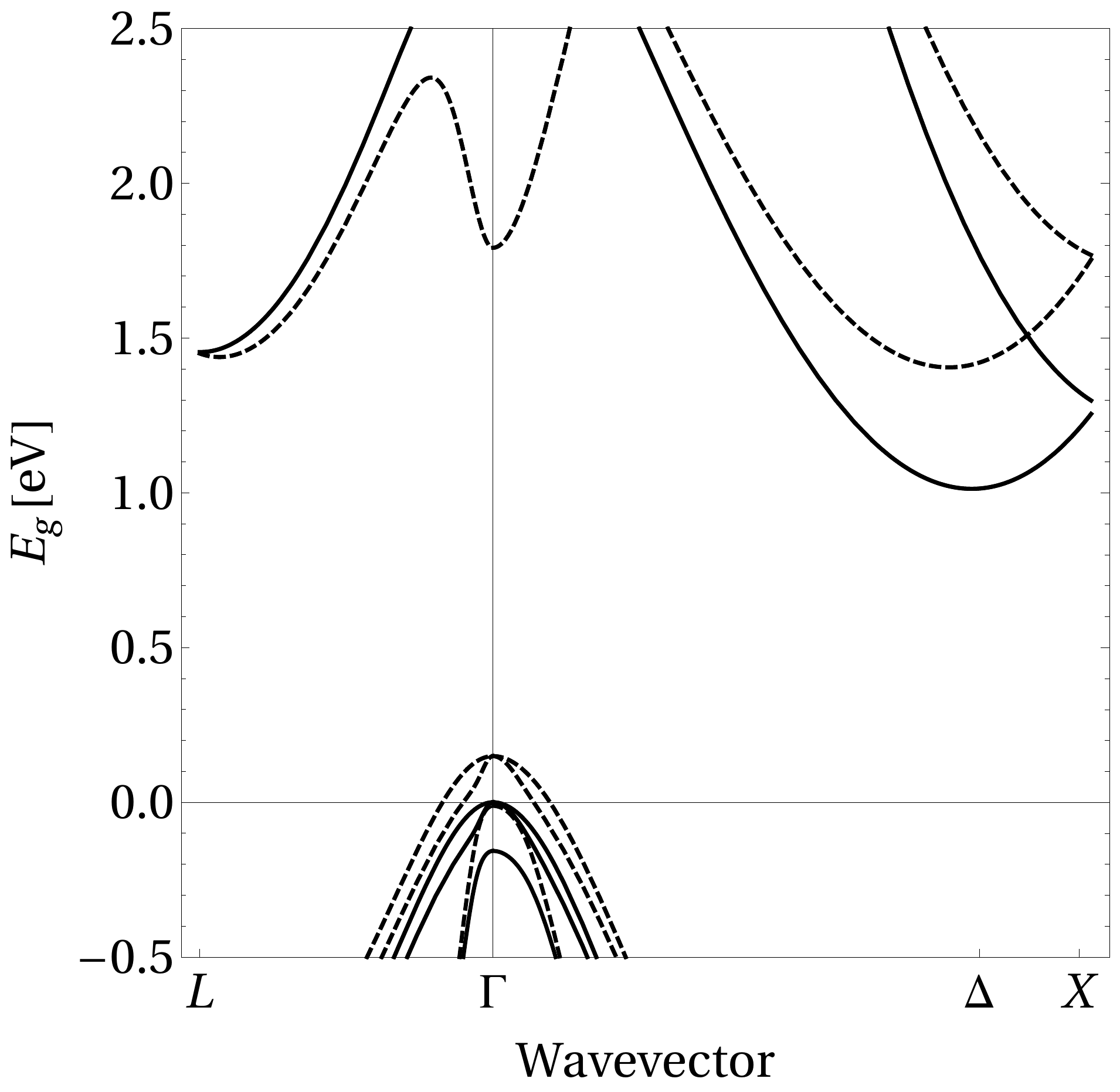}
\caption{\label{enatxp5} Zero temperature energy dispersion versus wavevector of SiGe with $x=0.5$. Solid and dashed lines represent strains of 0\% and 3\% respectively. The dispersions were calculated using the k.p approach of Ref. \onlinecite{kdp}.}
\end{figure}

\begin{figure}
\includegraphics[width=3.4in]{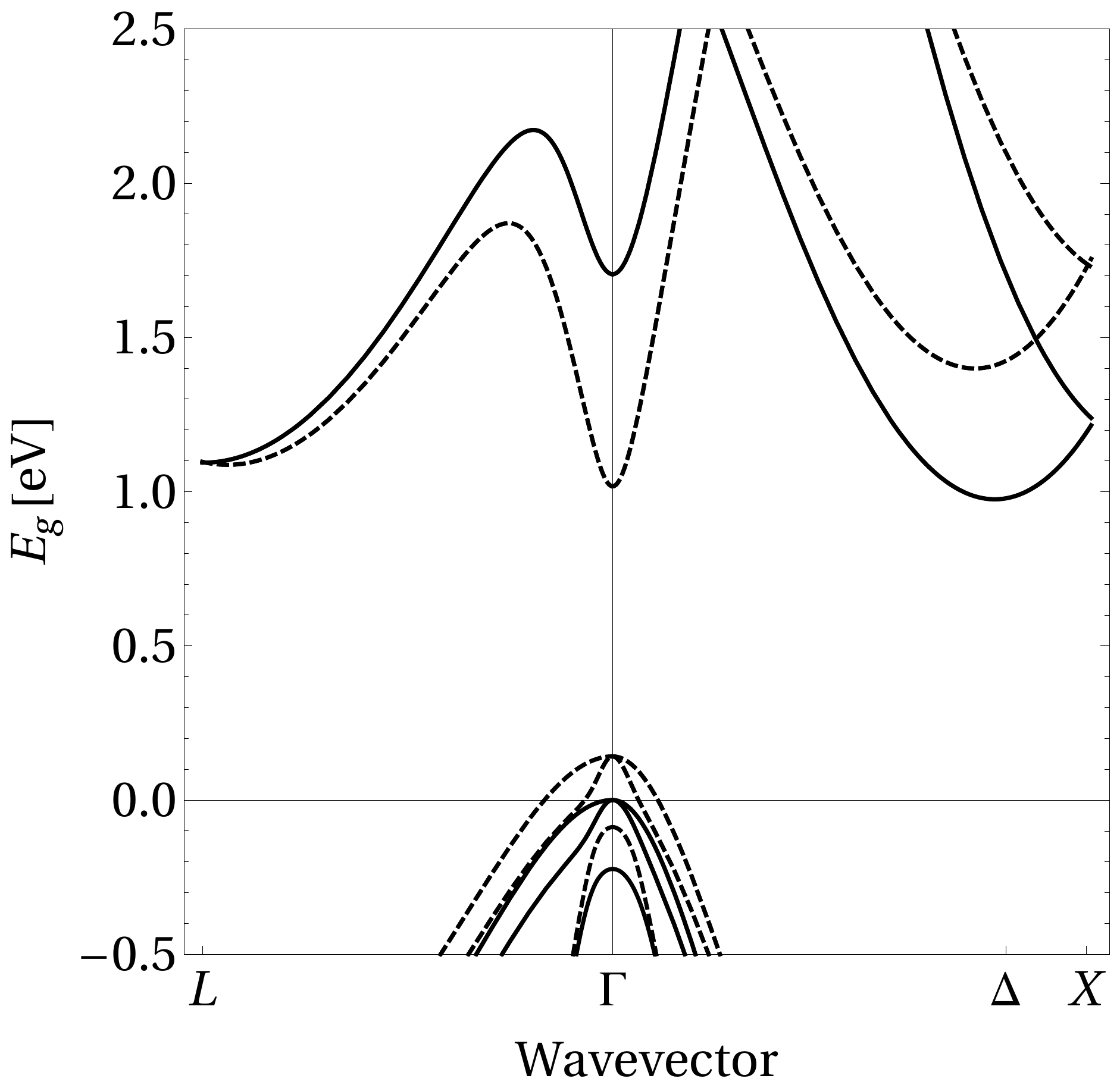}
\caption{\label{enatxp75} Zero temperature energy dispersion versus wavevector of SiGe with $x=0.75$. Solid and dashed lines represent strains of 0\% and 3\% respectively. The dispersions were calculated using the k.p approach of Ref. \onlinecite{kdp}.}
\end{figure}
\begin{figure}
\includegraphics[width=3.4in]{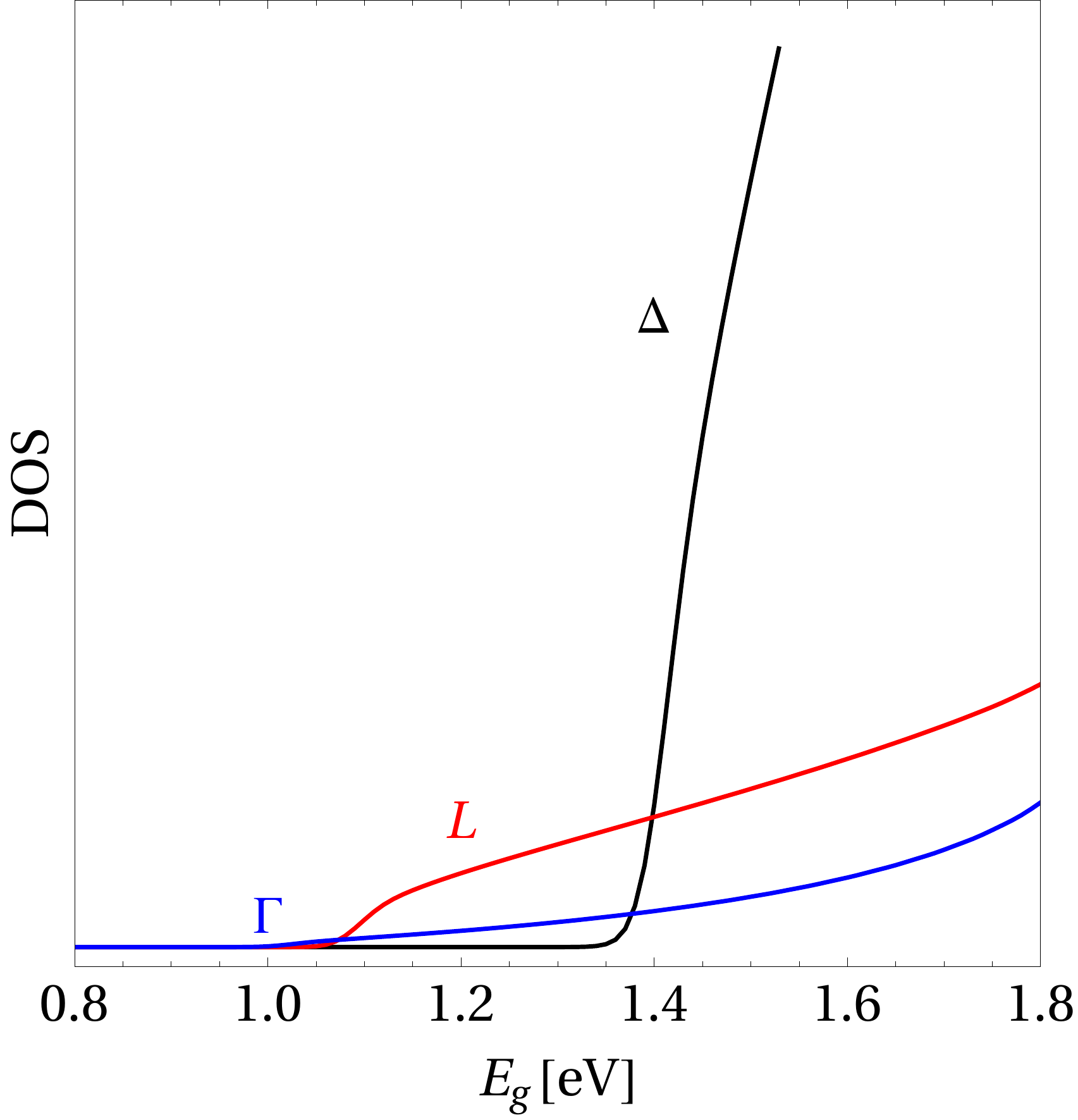}
\caption{\label{dosp75s3} Conduction band density of states of the $\Delta$, $L$ and $\Gamma$ valleys vs. energy of SiGe with $x=0.75$ at 3\% tensile hydrostatic strain.}
\end{figure}

\begin{figure*}
\includegraphics[width=6.8in]{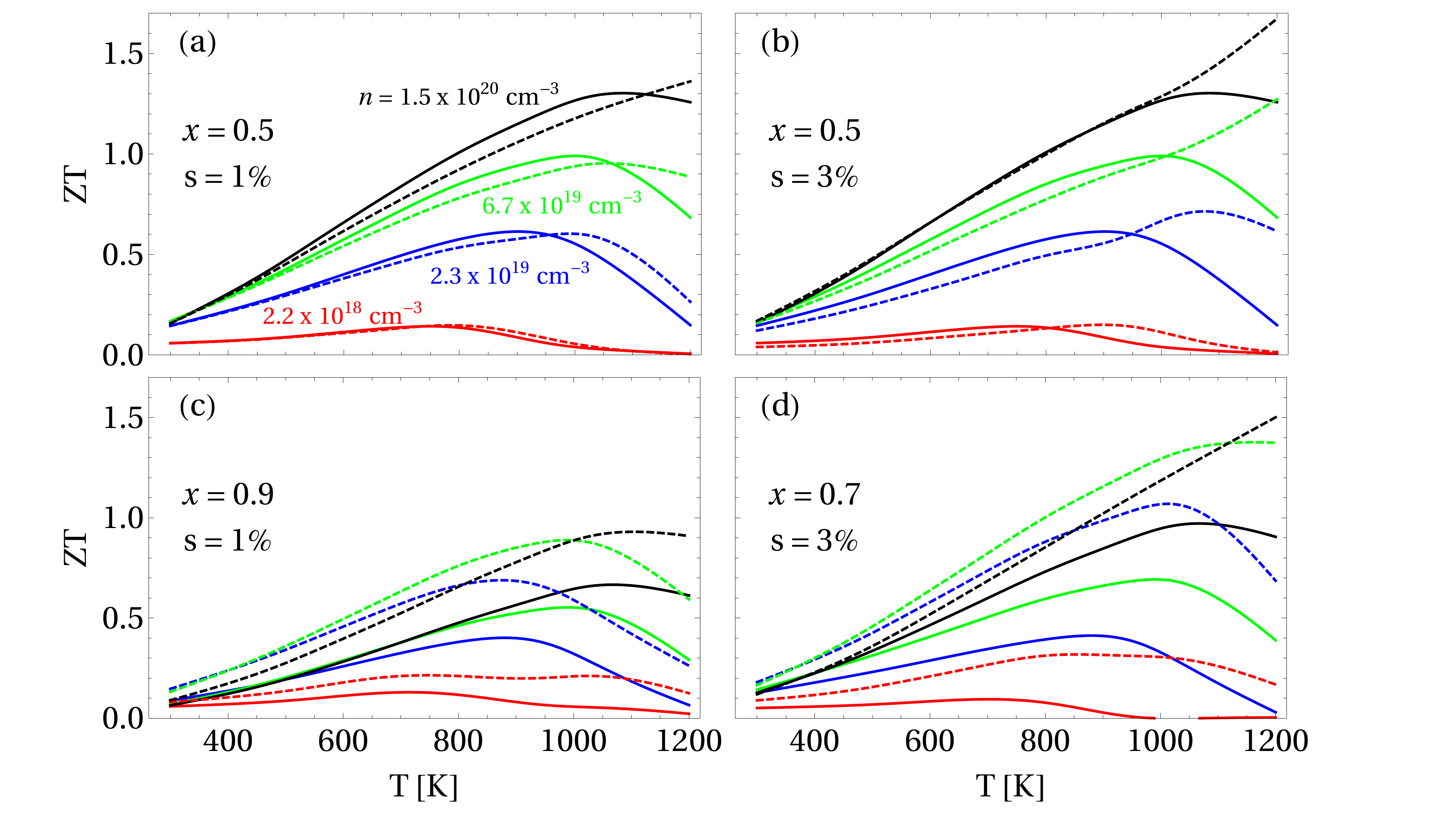}
\caption{\label{ztvstwnwst} Si$_{1-x}$Ge$_x$ $n$-type thermoelectric figure of merit $ZT$ vs temperature $T$ at four doping concentrations and (a) Ge composition $x=0.5$ and strain $s=1\%$ (b) $x=0.5$ and $s=3\%$, (c) $x=0.9$ and $s=1\%$ and (d) $x=0.7$ and  $s=3\%$. Solid and dashed lines represent the unstrained and hydrostatically tensile strained material, respectively.}
\end{figure*}

In this work, we illustrate the effect of strain on the thermoelectric properties of SiGe by applying tensile hydrostatic strain. We performed an exhaustive search in the phase space of Ge composition, strain, temperature and doping concentration, and found two interesting types of $ZT$ enhancements, shown in Fig. \ref{ztpfsigsiklvsxst} (a), achieved by 3\% tensile hydrostatic strain. This type of strain shifts the 6 $\Delta$ valleys up and the $\Gamma$ valley down in energy relative to the 4 $L$ valleys. This has the effect of moving the $\Delta-L$ valley cross-over towards the middle of the Ge compositions, at which the thermal conductivity is the lowest (see Fig. \ref{bandsn1p5d20} and band structures in Figs. \ref{enatxp5} and \ref{enatxp75}). This affects the figure of merit in three ways: i) having carriers in the higher mobility $L$ valley increases the overall conductivity at the central compositions, while the larger density of states at the $\Delta-L$ cross-over increases the Seebeck coefficient (see Fig. \ref{ztpfsigsiklvsxst} (c) and (d), and Fig. \ref{dosp75s3}), ii) the Lorenz number decreases, reducing the overall thermal conductivity (see Fig. \ref{ztpfsigsiklvsxst} (e) and (f)), and iii) the band gap is increased, thus pulling carriers out of the valence band. Consequently, the onset of bi-polar behaviour occurs at higher temperatures,
delaying the drop in Seebeck coefficient and $ZT$ to higher temperatures until the new onset of bi-polar behaviour (see Fig. \ref{ztpfsigsiklvsxst} (d) and \ref{ztvstwnwst} (b)). The suppression of the thermal conductivity via the reduction of the Lorenz number is only possible at high doping concentrations and temperatures, due to the large proportion of heat carried by the electrons (as seen in Fig. \ref{kratio}), where the electrons carry up to 50\% of the total heat. This type of hydrostatic strain may be achieved by a small amount of Sn added to the SiGe alloy.

The enhancement peak at Ge compositions $\sim 70\%$ is due to an increase in electronic conductivity due to the population of the higher mobility $\Gamma$ band\cite{jap}, lowered in energy by strain (see Fig. \ref{enatxp75}). The proximity of the $L$ valley retains a high density of states, avoiding a strong decrease of the Seebeck coefficient (see density of states at $x=0.75$ in Fig. \ref{dosp75s3}, corresponding to the band structure of Fig. \ref{enatxp75}). This results in an overall increase in Power Factor (PF=$S^2\sigma$). In this case, the increase in electrical conductivity increases the thermal conductivity via the Wiedemann-Franz law, and an optimum condition in Ge composition needs to be achieved to obtain an increase in $ZT$.

The two enhancements peaks in Fig. \ref{ztpfsigsiklvsxst} (a) behave very differently to each other with changing temperature and strain conditions, as seen in Fig. \ref{ztvstwnwst}. Hydrostatic tensile strain only enhances $ZT$ at 50\% Ge at very high temperatures, while at lower temperatures it has a detrimental effect. This is because only at high temperature the electrons carry enough of the heat for the reduction in the $\kappa_e$ to make a difference. Figures \ref{ztvstwnwst} (a) and (b) show the enhancement at $x=0.5$ through reduction in $\kappa_e$ at 1\% and 3\% tensile hydrostatic strain, respectively.

On the other hand, the $ZT$ increase at higher Ge composition is stable throughout the whole temperature range as seen in Figs. \ref{ztvstwnwst} (c) and (d). This enhancement is produced by a large increase in the conductivity, thanks to the population of the higher mobility $\Gamma$ valley, with a moderate decrease of the Seebeck coefficient resulting from the higher density of states $L$ valley. Interestingly, the optimum for this type of enhancement does not occur at the highest doping concentration, due to the higher base Seebeck coefficient at lower carrier concentrations.

When using strain to enhance $ZT$, the Ge composition at which the maximum of $ZT$ occurs is strain dependent. This can be inferred from the change in composition of the $\Delta-L$ and $\Gamma-L$ cross-overs at different strains in Fig. \ref{bandsn1p5d20}. Therefore, the composition of the alloy needs to be tailored to the strain to be applied, or vice versa. 

\section{Discussion}
From these results we can extract some general directions to improve $ZT$ using strain or changes in alloy composition. The first and most important is that increasing the band gap can achieve higher $ZT$ at higher temperatures by pushing the onset of bi-polar behaviour to a higher temperature. 

Secondly, we can increase the power factor via populating a high mobility band near a high density of states band, increasing $\sigma$ while preventing a large decrease in $S$. This can be achieved without much effect on the thermal conductivity, as in 3\% strained Si$_{0.3}$Ge$_{0.7}$. In this case, hydrostatic strain lowers the $\Gamma$ conduction band valley below the $L$ band. The $\Gamma$ has a low density of states and very high mobility.\cite{jap,fisch} The power factor enhancement is achieved near the composition of the $\Gamma-L$ cross over, at which the carriers populate the high mobility $L$ and higher mobility $\Gamma$ valleys, increasing the conductivity, while retaining a high density of states that prevents a too severe reduction of the Seebeck coefficient. 

Thirdly, in materials working at high temperature and/or very high doping concentration, reducing the Lorenz factor is an effective way to reduce the thermal conductivity, as most of the heat is carried by electrons. Significant increases in $ZT$ can be achieved if in addition the carriers populate a high mobility band with a high density of states, as is the case in 3\% strained Si$_{0.5}$Ge$_{0.5}$.
In this case the enhancement is achieved by positioning the $\Delta-L$ conduction band cross-over at Ge compositions with low thermal conductivity. The increased density of states forces the Fermi level into the gap away from the conduction band edge, enhancing Seebeck and reducing the electronic contribution to the thermal conductivity $\kappa_e$. The extra scattering reduces the conductivity somewhat, but not enough to reduce the power factor, due to the contribution from the high mobility $L$ band.

\section{Conclusion}
We have used first-principles electronic structure theory calculations to perform an exhaustive search for the highest enhancement of the thermoelectric figure of merit $ZT$ in the parameter space of $n$-type SiGe alloy composition, doping, temperature and hydrostatic strain. 
We found two promising configurations that illustrate two different and interesting approaches to improve $ZT$. Both are achieved by the application of 3\% tensile hydrostatic strain. We also obtain an excellent agreement between our calculated values and experiment for the mobility, Seebeck, electrical conductivity and thermoelectric figure of merit $ZT$.

The improvements on $ZT$ discussed need 3\% tensile hydrostatic strain on a $\sim$50\% Ge SiGe alloy. At this time, we know of no way to fully achieve this. A similar result may be achieved by alloying with a larger element, like Sn, or by growth on a substrate with a larger lattice constant. The latter is unlikely to work, since at strains over 1\% the critical thickness would become an issue, and only very thin layers could be grown. 
Alloying with Sn is more promising. The addition of 6\% Sn would produce strains of $\sim$1\%, and 18\% Sn would achieve 3\% strain in Si$_{0.5}$Ge$_{0.5}$.\cite{sigesn} We expect the additional disorder to have a small effect on the conductivity, as the SiGe is already maximally disordered. Incorporation of Sn on the order of 15\% has been achieved experimentally.\cite{solsigesn} However, the equilibrium solubility of Sn is much less than 18\% in SiGe, especially at the temperatures considered here.\cite{solsigesn} Nevertheless, there is hope for this material, as the thermal stability temperature of Si$_{0.17}$Ge$_{0.83}$Sn$_{0.15}$ has been found to be 450 $^{\circ}$C, and has been increasing with new growth techniques.\cite{solsigesn} Also, incorporation of Si in SiGeSn has been found to slow down phase separation.\cite{solsigesn}
A third option would be a large atomic dopant, but much further study is required to evaluate these options.

The main message of this work is to highlight the type of enhancements in $ZT$ that can be achieved by simultaneously populating electronic bands with different characteristics. As usual in this problem, the details matter, as changes in the electronic distribution (determined by the Fermi energy and temperature), and the population of the bands, can result in very different outcomes for the same electronic structure.

\section{Acknowledgement}
This work has been funded by Science Foundation Ireland award 12/IA/1601.

\bibliography{thermorprb}

\end{document}